\documentclass[aps,twocolumn,showpacs,floatfix]{revtex4}

\usepackage{epsfig}

\newcommand{\be}{\begin{equation}}
\newcommand{\ee}{\end{equation}}

\begin{document}

\title{Forces between a single atom and its distant mirror image
}

\author{Pavel Bushev, Alex Wilson, J\"urgen Eschner$^*$, Christoph Raab,
Ferdinand Schmidt-Kaler, Christoph Becher, and Rainer Blatt}

\affiliation{Institut f{\"u}r Experimentalphysik, Technikerstr.\ 25, A-6020
Innsbruck, Austria}

\date{\today}

\begin{abstract}

An excited-state atom whose emitted light is back-reflected by a
distant mirror can experience trapping forces, because the
presence of the mirror modifies both the electromagnetic vacuum
field and the atom's own radiation reaction field. We demonstrate
this mechanical action using a single trapped barium ion. We
observe the trapping conditions to be notably altered when the
distant mirror is shifted by an optical wavelength. The
well-localised barium ion enables the spatial dependence of the
forces to be measured explicitly. The experiment has implications
for quantum information processing and may be regarded as the most
elementary optical tweezers.

\end{abstract}


\maketitle


An atom which sits in the vicinity of mirrors or reflectors
experiences energy shifts of its electronic states. These level
shifts are known as the van der Waals, Casimir-Polder
\cite{Casimir1948} and resonant radiative shifts
\cite{Lyuboshitz1967, Morawitz1969}, the latter of which is caused
by a retarded interaction of the atom with its own radiation
field. For an atom in its excited state and at distances from a
mirror much less than the transition wavelength, the level shift
will be dominated by the van der Waals interaction, whilst in the
far field the level shift is attributed to the resonant
interaction with its own reflected field \cite{Morawitz1969,
Barton1987, Meschede1990, Hinds1991}. Such far-field shifts have
been observed with an atomic beam traversing an optical resonator
\cite{Heinzen1987} and with atoms in a microwave cavity
\cite{Brune1994}. The same effect has been predicted for a single
trapped ion whose emitted radiation field is reflected back by a
single, distant mirror \cite{Dorner2002}, and recently this level
shift has been observed with an indirect spectroscopic method
\cite{Wilson2003}.

The far-field mirror-induced shift of an excited atomic level
oscillates on the wavelength scale when the atom-mirror distance
is varied. Therefore, when the position of the atom is controlled
to the extent that it becomes sensitive to this spatial
dependence, then the level shift acts as a spatially varying
potential $U(\vec{r})$, and the atom feels its gradient
$-\vec{\nabla} U(\vec{r})$ as a force.

This mirror-induced force is a peculiar manifestation of the
mechanical effects of light. Forces due to applied light fields
were first demonstrated experimentally by Lebedev
\cite{Lebedev1901}, and the recoil of an absorbed photon on an
atom was observed by Frisch who deflected an atomic beam with
incoherent light \cite{Frisch1933}. With the advent of the laser,
such forces have found many important applications, from
decelerating, cooling and trapping atoms to optical tweezers in
biology \cite{Dholakia2002}.

%
Mirror-induced forces on individual atoms were first considered in
connection with cavity-QED experiments, where their use has been
proposed for trapping atoms in an optical resonator
\cite{Haroche1991, Schoen2003}.
It is this kind of binding force which we observe in the
experiment reported here. A single, trapped and laser-excited ion
is an ideal system for this observation, as its position can be
controlled on the nanometer scale \cite{Eschner2001,
Guthoehrlein2001, Mundt2002}, and interaction with a distant
mirror has already been demonstrated \cite{Eschner2001,
Kreuter2003, Wilson2003}.  These earlier experiments detected the
effect of a mirror on the internal, electronic state of an ion. In
contrast, our new study reveals directly the action on the ion's
motional degree of freedom, whereby we have a new level of control
over the total state of the atom and new possibilities for its
manipulation. While the previous observations \cite{Wilson2003}
have implications, e.g., for precision spectroscopy, our new
results are more relevant for studies of single ions in optical
cavities, for their cooling \cite{Domokos2003} and their
application in quantum information processing.

A mirror-induced energy shift of an excited state, like a modified
spontaneous decay rate \cite{Eschner2001, Kreuter2003}, has an
analogy in classical electrodynamics. The classical effect is
used, for example, to modify the emission diagram of an antenna by
reflectors, and it is well-known that such geometric modifications
also change the resonance frequency \cite{Blake}. These effects
can be treated in terms of radiation reaction only, i.e.\ in terms
of the interaction with the reflected field. The quantum
electrodynamic picture is quite different though, due to the
presence of the vacuum field which also couples to the atom
\cite{Milonni}. The concept of the vacuum field forms the
prevailing language in the field of experimental cavity QED, see
for example Refs.~\cite{Heinzen1987, Brune1994}. It was also used
in the proposals to trap an atom in a resonator \cite{Haroche1991,
Schoen2003} to which our experiment is closely related.
Rigorously speaking, however, the vacuum field alone cannot
account for spontaneous decay or its modification by reflectors,
but radiation reaction must also contribute \cite{Ackerhalt1973,
Milonni1973, Dalibard1982, Hinds1991a}. The same is true for
excited-state level shifts. In fact, the degree to which vacuum
fields and radiation reaction are seen to contribute depends upon
the ordering of operators in the Heisenberg equations of motion,
the choice of which has been called a "matter of taste"
\cite{Milonni}. In what follows we will use the concept of vacuum
fields as a convenient language but without insisting on any
particular distinction between vacuum fields and radiation
reaction.



In our experiment with a single trapped Ba$^+$ ion,
a fraction $\epsilon$ of
its fluorescence light is retro-reflected and focussed back upon
the emitting particle (see Fig.~\ref{setup} below). For this
situation, the model by Dorner \cite{Dorner2002} predicts an
energy shift of the excited level by
\be U(z) = -\hbar \frac {\epsilon \Gamma} {2} \sin (2 k z)~~.
\label{potential} \ee
Here $\Gamma$ is the decay rate
of the excited level, $k=2\pi/\lambda$ is the wave vector of the
light with wavelength $\lambda$, and $z$ is the position of the
mirror with respect to the ion. Depending on $z$, this potential
creates different mechanical effects: around $\sin(2kz)=0$, a
force is exerted on the ion which points either towards or away
from the mirror; around $\sin(2kz)=\pm 1$, a binding ($+1$) or
anti-binding ($-1$) potential is formed. Since the atom feels the
mirror-induced potential only whilst it resides in the excited
level, the forces are scaled by the probability $P_e$ for the atom
to be in that state. The maximum force at $\sin(2kz)=0$ is
therefore calculated as $P_e \hbar k\epsilon\Gamma$. The binding /
anti-binding potential at $\sin(2kz)=\pm 1$ is characterised by
the oscillation frequency which an otherwise force-free atom would
have in the respective potential well, $\omega_{vac}= (2P_e
\epsilon \Gamma \hbar k^2/m)^{1/2}$, with atomic mass $m$. For an
ion which is already confined with trap frequency $\omega_{trap}$
(typically around $2\pi \times 1$~MHz), the potential $U(z)$ of
Eq.~(\ref{potential}) adds to the trapping potential, thus
changing the trap frequency according to $\omega_{trap}^{\prime} =
(\omega_{trap}^2 + \omega_{vac}^2 \sin(2kz))^{1/2}$. Since the
deviation $\delta\omega_{trap} = \omega_{trap}^{\prime} -
\omega_{trap}$ is small, it is well approximated by
\be \delta \omega_{trap}(z) = \frac{P_e \epsilon \Gamma \hbar k^2}
{m \omega_{trap}} \sin(2kz)~~. \label{freqshift} \ee
It is this change of the trap frequency, a direct mechanical
action, which we measure in the experiment. We emphasize that the
level shift $U(z)$ producing this extra trapping force is caused
by the presence of a single distant mirror and the associated
modification of vacuum and radiation reaction fields. This
distinguishes our observations from the recently demonstrated
trapping of atoms in excited high-finesse optical resonators,
where the mechanical action arises not only from reradiated
photons, but also from the externally excited resonator mode
\cite{Pinkse2000, Hood2000}.



\begin{figure}[tb]
\epsfig{file=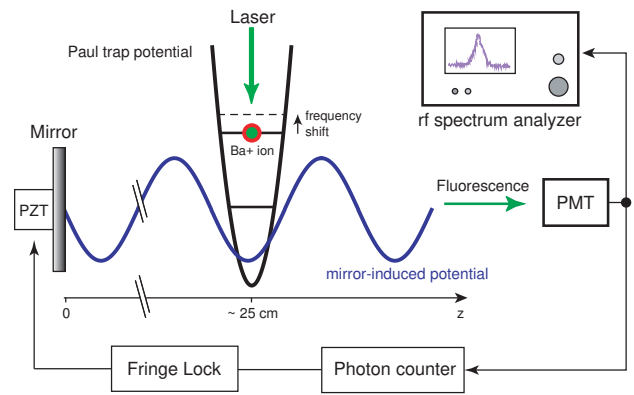, width=0.95\hsize} 
\caption{\label{setup} Principle of the experiment. A single
trapped $^{138}$Ba$^+$ ion is laser-excited at 493~nm. A
retro-reflecting mirror 25~cm away from the trap and a lens (not
shown) are arranged such that they image the ion onto itself. The
493~nm fluorescence is detected by a photomultiplier (PMT). Its
intensity modulation, i.e.\ the motional sideband due to the ion's
oscillation in the trap, is recorded with a spectrum analyser.
From the three different trap vibrations, we observe the one at
the lowest frequency, $\omega_x \approx 2\pi \times 1.02$~MHz,
whose orientation is at about 54$^{\circ}$ to the optical axis.
The deviation of the mean count rate from a chosen offset value is
used in a feedback loop to control the position of the mirror such
that the ion stays at a given point on an interference fringe to
within $\sim 10$~nm. In the diagram, PZT stands for
piezomechanical translator. The feedback loop has an integration
time of about 1~s and compensates for slow drifts of the
ion-mirror distance but not for the ion's oscillation in the trap.
By switching the sign of the feedback gain, we choose between the
positive and negative slopes of the interference fringes. More
details of the setup are found in Refs.~\cite{Eschner2001,
Wilson2003}.}
\end{figure}

The experimental setup is shown in Fig.~\ref{setup}. The counting
signal on the PMT exhibits high-contrast interference fringes as
the ion-mirror distance is varied \cite{Eschner2001}. This
interference signal follows the $-\cos(2kz)$ dependence of the
modified 493~nm decay rate \cite{Dorner2002}, such that the
midpoints of the slopes correspond to $\sin(2kz)=\pm1$, i.e.\ to
the maximum binding or anti-binding potential, as described above
(see also Fig.~\ref{spatial} below). The trap frequency is
measured by spectrally analysing the PMT signal. It contains a
spectral component at the trap frequency, around 1~MHz, because
the oscillation of the ion creates an intensity modulation of the
scattered light. The signal on the spectrum analyser has, to good
approximation, a Lorentzian line shape with width $\Delta f$ of
about 500~Hz. After a few seconds of averaging, the centre
frequency of the line is determined with less than 10~Hz
inaccuracy.

\begin{figure}[htb]
\epsfig{file=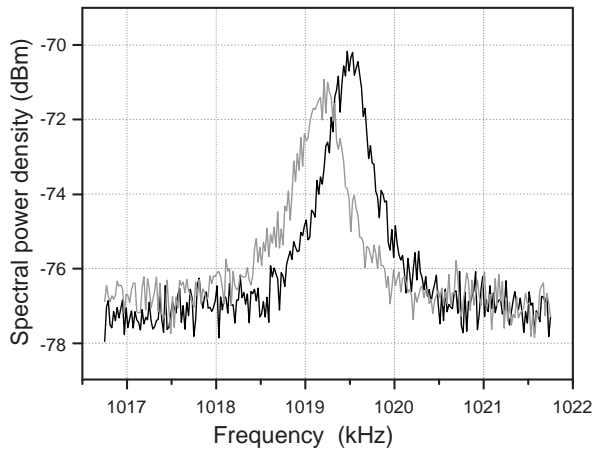, width=0.9\hsize} 
\caption{\label{Lorentzians} Signal on the spectrum analyser for
the ion positioned on the positive (right curve) and negative
slope (left curve) of the interference signal. The centre
frequency of a Lorentzian fit to the data is taken as the trap
frequency. The broadening of the lines is a consequence of the
ongoing laser cooling of the ion, and the width represents the
steady-state cooling and heating rate \protect{\cite{Cirac1995a}}.
The measured values agrees well with the expectation
\protect{\cite{Raab2000}}. The size of the Lorentzian curve above
the Poissonian noise level is observed to vary between 2 and 10~dB
and serves as a measure of the amplitude of the ion's oscillation
in the trap. }
\end{figure}

Fig.~\ref{Lorentzians} shows two spectra which were recorded
directly one after the other, with the ion positioned on the
midpoints of a positive and negative slope of the interference
signal, respectively. The shift is clearly visible and amounts to
310~Hz in this case. The value is within the range expected from
Eq.~\ref{freqshift}, which predicts around 350~Hz, taking typical
values for our experiment $P_e \approx 7\%$, $\epsilon \approx
1.5\%$, $\lambda = 493$~nm, and $\Gamma = 2 \pi \times 15.4$~MHz.
The corresponding value of $\omega_{vac}$ is about 20~kHz, larger
than the photon recoil frequency of 6~kHz. We emphasize that no
changes are made to the setup between the recording of the two
spectra, apart from translating the distant mirror by $\lambda/4$.

\begin{figure}[tb]
\epsfig{file=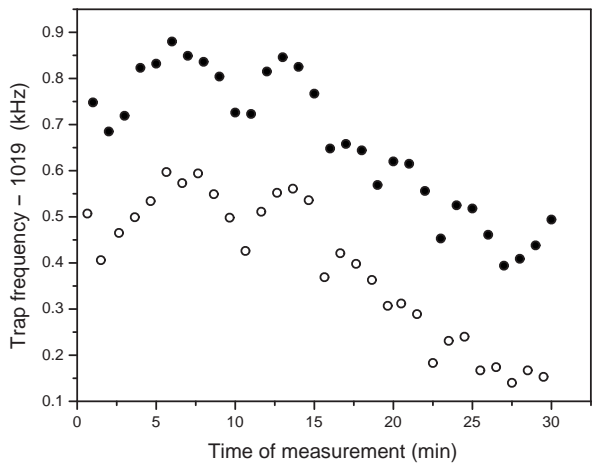, width=0.9\hsize} 
\caption{\label{centerfreq} Trap frequency measured on the positive (full
circles) and negative slope (open circles) of the interference signal, vs.
measurement time. }
\end{figure}

To measure the value of the frequency shift for a particular set of parameters,
we record about 60 spectra, alternating between the two slopes. Each spectrum
is fitted by a Lorentzian, and the centre frequency is plotted. An example is
shown in Fig.~\ref{centerfreq}. While the trap frequency itself varies due to
slow drifts of the trap drive intensity and due to thermal effects, a constant
difference is observed between the values measured on the two slopes. The
precise value of the shift depends on details of the experiment such as the
settings of the lasers, their directions, and the fine alignment of the
back-reflecting mirror. We observe values between 50 and 350~Hz, all within the
range expected from Eq.~\ref{freqshift}. It is important to note that we always
find the higher trap frequency on the positive slope of the interference
fringes (count rate vs. ion-mirror distance), in agreement with the theoretical
prediction \cite{Dorner2002}.

\begin{figure}[htb]
\epsfig{file=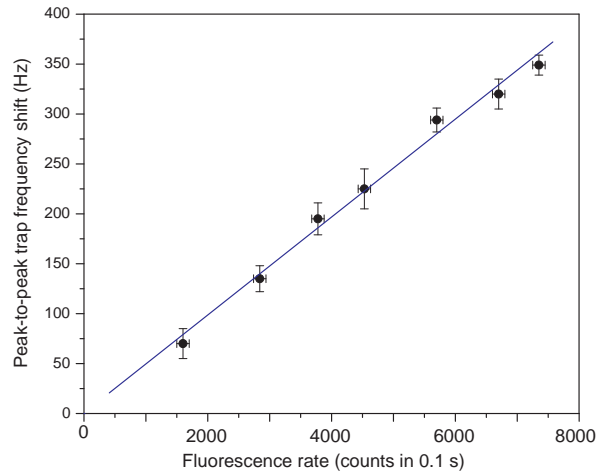, width=0.9\hsize} 
\caption{\label{powerdependence} Measured dependence of trap frequency
variation on excited state population $P_e$. The peak-to-peak difference,
between the midpoints of the two slopes of the interference fringes, is plotted
vs.\ the mean count rate. 10000 counts correspond to $P_e \approx 0.1$. The
line is a linear fit.
}
\end{figure}

As shown in Eq.~\ref{freqshift}, the trap frequency shift depends on the laser
parameters through the probability $P_e$ with which the ion is found in the
excited state. This dependence has been measured by recording the maximum shift
for different laser parameters. The mean fluorescence level, at the midpoint of
the interference fringes, serves as an indicator of $P_e$, to which it is
strictly proportional. The result is displayed in Fig.~\ref{powerdependence}.
The data agree well with the expected linear dependence. A further test is the
dependence of the trap frequency on the position of the mirror. When we shift
the ion between the maxima and minima of the interference fringes, we find the
result shown in Fig.~\ref{spatial}. The sinusoidal variation predicted by
Eq.~(\ref{freqshift}) is clearly observed.

\begin{figure}[htb]
\epsfig{file=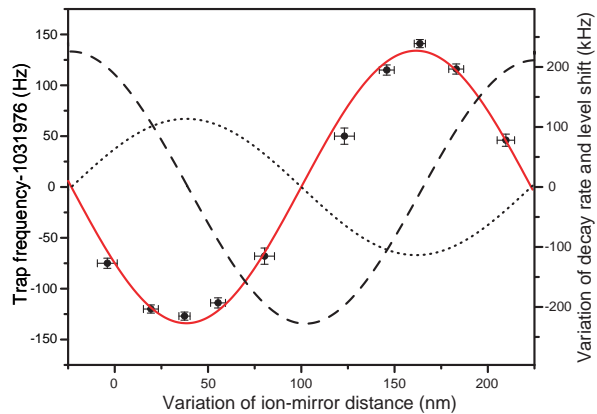, width=0.9\hsize} 
\caption{\label{spatial} Measured variation of the trap frequency
with ion-mirror distance (data points). The distance is adjusted
by varying the offset level in the feedback loop to the PZT. This
data set showed particularly small drifts of the trap frequency.
The solid line is a fit to the data. The dashed line shows the
corresponding, calculated variation of the spontaneous decay rate
on the S-P transition, and the dotted line is the shift of the
excited level, i.e.\ the vacuum potential divided by $\hbar$. The
dashed and dotted lines use the right-hand vertical scale. The
calculated maximum force, acting when the atom is positioned on a
maximum or minimum of the dashed curve, corresponds to an
acceleration of $\sim 100~g$.}
\end{figure}

One may construct a semi-classical explanation for the observed
mechanical action, analogous to the intuitive picture in
\cite{Dorner2002} that the level shift corresponds to the energy
of the atomic dipole in the light field returning from the mirror.
When a maximum in the interference fringes is observed at the PMT,
the returning light stimulates additional radiation towards the
PMT, thus creating a small recoil towards the mirror. Conversely,
a minimum in the interference fringes corresponds to radiation
returning from the mirror being predominantly absorbed, which
therefore leads to a force away from the mirror. However, the
quantum mechanical properties of the electromagnetic field are
needed to explain quantitatively the spontaneous emission rate
from an atom \cite{Heitler, Milonni, Hinds1991a}. Therefore the
concepts of vacuum fields and radiation reaction, as presented in
the introduction, are felt to be the most accurate way of
describing our experimental findings.

We emphasize that it is a distant and passive optical element
which introduces a controlled, position-dependent mechanical
action on the atom in our experiment. In general, any dielectric
optical element which back-reflects the light scattered from
nearby atoms shifts the excited levels, modifies the transition
frequencies, and acts on the motional state. Therefore our study
is also relevant for technological applications of single trapped
atoms and ions, in particular when they are combined with
high-finesse optical cavities.
On the other hand, since energy shifts of individual levels
accumulate in time to phase shifts of the atomic wavefunction,
their control, as demonstrated here, may become useful for the
manipulation of optical phases in applications of single atoms or
ions for quantum state engineering or quantum information
processing. Another possible application would be "vacuum optical
tweezers" which exert forces on a laser-excited molecule just by
an arrangement of microscopic mirrors.

\vspace{12pt} \noindent \textbf{Acknowledgements.} This work is
supported by the Austrian Science Fund (FWF, SFB15), the European
Commission (QUEST network, HPRNCT\--2000\--00121, QUBITS network,
IST\--1999\--13021), and the "Institut f\"ur Quanteninformation
GmbH". P.~B.\ thanks A.~Allahverdyan and K.~Lamonova for
stimulating discussions. The authors thank Uwe Dorner for
clarifying remarks.

\end{document}